\documentclass[twocolumn,showpacs,a4paper]{revtex4}
\usepackage{graphicx}

\begin{document}

\title{Phase coherence and fragmentation in weakly interacting bosonic gases}

\author{G.~S.~Paraoanu}\email{paraoanu@cc.hut.fi}

\affiliation{Low Temperature Laboratory, Helsinki University of Technology,
P.O. Box 5100, FIN-02015 TKK, Finland}


\begin{abstract}
We present a theory of measurement-induced interference for weakly interacting Bose-Einstein condensed (BEC)
gases. The many-body state resulting from the evolution of an initial fragmented (Fock) state can be approximated
as a continuous superposition of Gross-Pitaevskii (GP) states; the
measurement breaks the initial phase symmetry, producing a distribution pattern corresponding to only one of the
GP solutions. We discuss also analytically solvable models, such as two-mode on-chip adiabatic recombination and
soliton generation in quasi one-dimensional condensates.

\end{abstract}

\maketitle

A long-standing fundamental problem in theoretical physics is to understand how relative phases
are established between superfluids that have never been in contact with each other \cite{anderson}. In the case of 
atomic Bose-Einstein condensates, the first experiment of the sort was done already a decade ago \cite{andrews}; 
strangely though, the data could be reproduced simply by assuming a phase relation between the initial condensates 
and using the time-dependent GP equation \cite{walls}. It soon became acknowledged, following a 
number of elegant proofs for the noninteracting case \cite{javanainen,ycastin}, that the U(1) phase symmetry
is broken by the measurement process itself. Since at the time of the measurement the density of the expanding 
condensates is relatively low, it is assumed that the situation is for all practical purposes equivalent to the 
noninteracting case. Since \cite{andrews}, better-controlled experiments in various setups (optical 
lattices \cite{dalibard}, on-chip split condensates \cite{ketterlenew}, nondestructive measurements 
\cite{pritchard}) have supported this view.

However, in the interacting case the situation has proven to be more complicated: on one hand, it is known that 
interaction within each cloud tends to delocalize the phase between two condensates, leading to squeezing and to 
other effects similar to the Josephson effect in superconductors \cite{papers}. But BEC interference experiments 
require overlap of the atomic clouds, therefore the physics at work might well be different. If atoms originating 
from the initially independent clouds interact, it is not even clear in what sense, at the measurement time, we can 
talk about condensates that have not seen each other. Since fragmented states are not robust to certain classes of 
perturbations, it could well happen that the interaction itself would lead to phase localization \cite{stringari}. 
Moreover, for repulsive interactions a typically phase-coherent ground state is energetically favored 
\cite{papers,mueller}. Very recently, some authors have calculated, using standard many-body techniques 
\cite{wrong} the average of the density operator on evolving fragmented states and found out that in the 
interacting case the average of the density operator could display ripples which look somewhat similar to the 
fringes observed in atomic interference experiments.
It is yet not clear if a phase-coherent condensate is formed due to interactions before the measurement starts 
\cite{wrong}. Could it be then the case that what is detected in BEC interference experiments with interaction is 
in fact the density and not higher-order correlations?

In this paper, we present a theoretical approach that takes into account both the effects of interaction and that of measurement. 
We show that, also for weakly interacting bosons, the broken-symmetry approach to this problem \cite{walls} is 
justifiable, and there is no contradiction with results \cite{wrong} showing ripples in the average particle 
density.

The Hamiltonian of a system of bosons with repulsive two-body interaction is
\begin{equation}
\hat{H} = \int
d{\bf r}\hat{\psi}^{\dagger}({\bf r})H_{0}\hat{\psi} ({\bf r}) + \frac{g}{2}\int
d{\bf r}\hat{\psi}^{\dagger}({\bf r})\hat{\psi}^{\dagger}({\bf r})\hat{\psi} ({\bf r})\hat{\psi}
({\bf r}), \label{hh}
\end{equation}
with $g= 4\pi\hbar^{2}a/m > 0$, and $H_{0}=(-\hbar^{2}/2m)\nabla^{2} + V_{\rm ext}({\bf r})$ the Hamiltonian of a
single particle moving in an external potential $V_{\rm ext}$. We start by presenting the general argument
for evolution and measurement; to gain further insight, we examine later in the paper particular cases in which
the GP equation can be solved analytically.

{\it Evolution} We first show that, starting from an initial fragmented state $\mid {\rm gnd}_{L} \rangle \mid {\rm gnd}_{R} \rangle$,
 the time-dependent many-body state can be approximated as
 \begin{equation}
 \exp[-\frac{i}{\hbar}\hat{H}t] \mid {\rm gnd}_{L} \rangle \mid {\rm gnd}_{R} \rangle
 \approx
 c_{0}\int_{0}^{2\pi}\frac{d\varphi}{2\pi}\mid \Phi_{\varphi }(t)\rangle_{N}, \label{state}
\end{equation}
where $c_{0} = 2^{N/2}(N/2)!/\sqrt{N!}$ \cite{ycastin}, $\mid\Phi_{\varphi }(t)\rangle_{N} = (\sqrt{N!})^{-1}\left[\int 
d^{3}{\bf r}\Phi_{\varphi }({\bf r},t)\hat{\psi}^{\dagger}({\bf r})\right]^{N}\mid 0\rangle$, and $\Phi_{\varphi }({\bf r},t)$
will be identified as
the result of evolving the initial state $\Phi_{\varphi }({\bf r},0) =
(1/\sqrt{2})\left(\phi_{L}({\bf r})e^{-i\varphi
/2}+\phi_{R}({\bf r})e^{i\varphi /2}\right)$ by the time-dependent GP equation. We will proceed in a
systematic way, using a perturbative expansion technique developed in a different context \cite{fetter,castin-gardiner}.
We first construct, at any time $t$ and for a given phase $\varphi$, a projection operator $R_{\varphi }(t)$ outside
the subspace spanned by the ket $\mid\Phi_{\varphi }(t)\rangle$, $R_{\varphi }(t)= 
1-\mid\Phi_{\varphi}(t)\rangle\langle\Phi_{\varphi}
(t)\mid $, and we split the field operator $\hat{\psi} ({\bf r}')=\hat{\Phi}_{\varphi}({\bf r}',t) + \hat{\chi}_{\varphi}
({\bf r}',t)$, where
\begin{equation} \hat{\chi}_{\varphi} ({\bf r}',t) = \int d{\bf r}
R_{\varphi}({\bf r}',{\bf r};t)\hat{\psi}({\bf r}).
\end{equation}
Next, we linearize the Hamiltonian (\ref{hh}) around $\hat{\Phi}_{\varphi}$
and, after relatively lengthy but straightforward calculations \cite{castin-gardiner}, we find that in the Heisenberg picture (denoted by 
a superscript $(\cal{H})$), the evolution of the
phonon operator $\hat{\chi}^{(\cal{H})}({\bf r}',t)$ takes the form
\begin{eqnarray}
i\hbar\frac{\partial}{\partial t}\hat{\chi}_{\varphi}^{(\cal{H})}({\bf r}',t) = \int
d{\bf r}R_{\varphi}({\bf r},{\bf r};t)\left[ -i\hbar\frac{\partial}{\partial
t}\hat{\Phi}_{\varphi}({\bf r};t) \right. && \nonumber \\
 \left.+ H_{0}\hat{\Phi}_{\varphi}({\bf r},t) +
g\hat{\Phi}^{\dagger}_{\varphi}({\bf r},t)\hat{\Phi}_{\varphi}({\bf r},t)\hat{\Phi}
_{\varphi}({\bf r},t)\right]^{(\cal{H})} ,&&
\end{eqnarray}
and therefore
\begin{widetext}
\begin{eqnarray}
i\hbar \frac{d}{dt}\int d\varphi ' _{N-1}\langle \Phi_{\varphi ´}(t)\mid \hat{\chi}_{\varphi}({\bf r}',t) \mid 
\Phi_{\varphi '}(t)\rangle_{N} =
\int d\varphi ' _{N-1}\langle \Phi_{\varphi ´}(0)\vert i\hbar \frac{d}{dt}
\hat{\chi}^{(\cal{H})}_{\varphi}({\bf r}',t) \vert \Phi_{\varphi '}(0)\rangle_{N} \nonumber \\
\simeq \sqrt{2\pi}\mid \frac{d^{2}}{d\varphi '^{2}}
\vert \langle \Phi_{\varphi '}\mid \Phi_{\varphi} \rangle (t)\mid_{\varphi ' = \varphi}
\vert^{-1/2}\int d{\bf r}R_{\varphi}({\bf r}',{\bf r};t)[
-i\hbar\frac{\partial}{\partial t}\Phi_{\varphi}({\bf r},t)
+ H_{0}\Phi_{\varphi}({\bf r},t) +
gN\mid\Phi_{\varphi}({\bf r},t)\mid^{2}\Phi_{\varphi}({\bf r},t)] \label{vraiste}
\end{eqnarray}
\end{widetext}
The last equality is written up to an (irrelevant) phase factor. To obtain it, we notice that, as an immediate 
application of the Cauchy-Bunyakovski-Schwarz inequality in $L^{2}$ Banach spaces,
$\mid \langle \Phi_{\varphi '}(t)\mid \Phi_{\varphi} (t)\rangle \mid$ reaches a maximum value of 1 for $\varphi ' = 
\varphi$.
The quantity
$\mid _{N}\langle \Phi_{\varphi '}(t)\mid \Phi_{\varphi} (t)\rangle_{N} \mid$
will then be very close to zero for $\varphi ' \neq \varphi$, and strongly peaked
to 1 if   $\varphi ' = \varphi$, so we can write

\begin{eqnarray}
&& \mid ~_{N}\langle \Phi_{\varphi '}(t)\mid \Phi_{\varphi} (t)\rangle_{N} \mid
\approx \nonumber \\
&& \approx 1 + \frac{N}{2}\frac{d^{2}}{d\varphi '^{2}}
\mid \langle \Phi_{\varphi '}\mid \Phi_{\varphi} \rangle (t)\mid_{\varphi ' = \varphi}
(\varphi ' -\varphi )^{2} \nonumber \\
&&
\approx \exp \left[-\frac{N}{2}\mid\frac{d^2}{d\varphi '^{2}}\mid\langle \Phi_{\varphi '}\mid \Phi_{\varphi}\rangle
(t)\mid_{\varphi ' = \varphi}\mid (\varphi '-\varphi )^2\right] \nonumber \\
&& \approx \sqrt\frac{2\pi}{N}\mid \frac{d^{2}}{d\varphi '^{2}}
\mid \langle \Phi_{\varphi '}\mid \Phi_{\varphi} \rangle (t)\mid_{\varphi ' = \varphi}
\mid^{-1/2}\delta (\varphi ' -\varphi) \nonumber
\end{eqnarray}

The selfconsistency condition for the validity of the expansion Eq. (\ref{state}) is that at any time the states 
${\Phi}_{\varphi}({\bf r},t)$
remain macroscopically occupied,
\begin{eqnarray}
\int d\varphi ' _{N-1}\langle \Phi_{\varphi '}(t)\mid \hat{\Phi}_{\varphi}({\bf r}',t)\mid\Phi_{\varphi }(t)\rangle_{N} && 
\\ \gg
\int d\varphi '_{N-1}\langle\Phi_{\varphi '}(t)\mid\hat{\chi}_{\varphi} ({\bf r}',t)\mid \Phi_{\varphi }(t)\rangle_{N}&& 
\label{condition}
\end{eqnarray}
or in other words the quantum depletion of a state $\varphi$ should not aquire contributions of the order of the 
corresponding order parameter during evolution. According to Eq. (\ref{vraiste}), this is realized if
$\Phi_{\varphi }({\bf r},t)$ is a solution of the time-dependent GP equation, $i\hbar (\partial /\partial
t)\Phi_{\varphi }
({\bf r},t) = H_{0}\Phi_{\varphi }({\bf r},t) +
gN\mid\Phi_{\varphi }
({\bf r},t)\mid^{2}\Phi_{\varphi } ({\bf r},t)$ with initial condition
$\Phi_{\varphi }({\bf r},0) =
\frac{1}{\sqrt{2}}\left(\phi_{L}({\bf r})e^{-i\varphi
/2}+\phi_{R}({\bf r})e^{i\varphi /2}\right)$.
This is because the evolution Hamiltonian
has vanishigly small matrix elements between states $\mid\Phi_{\varphi }(t)\rangle_{N}$ with different phases 
$\varphi$; therefore each of these states is evolved independently. The limits of validity of this formalism  are 
identical to those of the time-dependent Gross-Pitaevskii equation for a macroscopic occupation
of $\Phi_{\varphi}$, namely that dynamical instabilities do not develop (the Bogoliubov-de Gennes equations 
should not have imaginary eigenvalues). We also note that, using the orthonormality property described above, the 
same result can be obtained by minimizing the
action $\int_{0}^{\tau}dt [(1/2)\langle i\hbar (d /dt )\rangle + c.c. - \langle H \rangle]$ on the state 
Eq.(\ref{state}).

{\it Measurement} Suppose now that we have a series of $n\gg 1$ detections leading 
to coordinates ${\bf r}_{1}, {\bf r}_{2}, ..., {\bf r}_{n}$; the resulting wavefunction is then
\begin{equation}
\int d\varphi \prod_{j=1}^{n} \Phi_{\varphi }({\bf r}_{j},t)\mid \Phi_{\varphi }(t)\rangle_{N-n}.
\end{equation}
Given a certain sequence  ${\bf r}_{1}, {\bf r}_{2}, ..., {\bf r}_{n}$, there are two possibilities: either there is a wavefunction 
$\Phi_{\varphi }({\bf r},t)$ whose shape it approximates best - and in this case this wavefunction is selected -
or there is no such wavefunction and in this case the probability of this particular sequence is very small. To 
prove this statement, we proceed by mathematical induction.
Let us divide the detection space into small bins of volume $v$ each, centered at discrete positions ${\bf r}_{i}$. A 
particle can be detected in either one of the bins $i$. Suppose that
after $n$ detections we have $n_{i}$ atoms at ${\bf r}_{i}$, resulting in the wavefunction
\begin{equation}
\int d\varphi \prod_{i}\left[\Phi_{\varphi }({\bf r}_{j},t)\right]^{n_i}\mid \Phi_{\varphi }(t)\rangle_{N-n}.
\end{equation}
We now show that the maximum absolute value of the coefficient $\prod_{i} \left[\Phi_{\varphi 
}({\bf r}_{i},t)\right]^{n_i}$ corresponds to a $\tilde{\varphi}$ for which
$n_{i}/(n v) = \mid \Phi_{\tilde{\varphi} }({\bf r}_{i},t)\mid ^2$, in other words the
"histogram" of $n_i$'s is mimicked by the Born rule for the wavefunction corresponding to
$\tilde{\varphi}$. We use the method of Lagrange multipliers, with the normalization $\sum_{i}\mid \Phi_{\varphi 
}({\bf r}_{i},t)\mid^{2}v =1$ as a constraint; we then have
\[
\frac{\partial}{\partial\varphi}\left[\prod_{i}\mid  \Phi_{\varphi
}({\bf r}_{i},t)\mid^{2n_i}-\lambda\sum_{i}\mid \Phi_{\varphi }({\bf r}_{i},t)\mid^{2}v
\right] = 0 .
\]
A solution of this equation is obtained for $\mid \Phi_{\tilde{\varphi}}({\bf r}_{i},t)\mid^{2}/\mid \Phi_{\tilde{\varphi 
}}({\bf r}_{j},t)\mid^{2} = n_{i}/n_{j}$; supplemented with the normalization condition, this produces the announced 
result $n_{i} = nv\mid \Phi_{\tilde{\varphi}}({\bf r}_{i},t)\mid ^2$.

This means that the next measurement event
will tend to reduce even more the probability amplitude of states which differ significantly from $\mid  
\Phi_{\tilde{\varphi}}({\bf r}_{i},t)\mid$, since these states have anyway a smaller probability amplitude to start 
with.

We will now show that, as $n$ increases, the probability amplitudes $\prod_{i}\mid  \Phi_{\varphi 
}({\bf r}_{i},t)\mid^{n_i}$ peak strongly around a value corresponding to
$\tilde{\varphi}$. We make a Taylor expansion around the maximum value,
\[
\prod_{i}\mid  \Phi_{\varphi }({\bf r}_{i},t)\mid^{n_i}\approx \prod_{i}\mid  \Phi_{\tilde{\varphi}}
({\bf r}_{i},t)\mid^{n_i} + \frac{1}{2} [Y]^t {\cal M} [Y] \nonumber ,
\]
where $[Y]$ is a column vector with transpose $[Y]^t = ( ...,\Phi_{\varphi }({\bf r}_{i},t) - \Phi_{\tilde{\varphi}}
({\bf r}_{i},t) , ... )$, and ${\cal M}$ is the Hessian matrix
with elements ${\cal M}_{ii} = (n_{i} -1)nv\prod_{i}\mid  \Phi_{\tilde{\varphi}}({\bf r}_{i},t)\mid^{n_i}$ and ${\cal 
M}_{ij} = \sqrt{n_{i}n_{j}}\prod_{i}\mid  \Phi_{\tilde{\varphi}}({\bf r}_{i},t)\mid^{n_i}$ ($i\neq j$). Using the 
constraint imposed by normalization, we find
\begin{eqnarray}
&&\prod_{i}\mid  \Phi_{\varphi }({\bf r}_{i},t)\mid^{n_i}\approx \frac{\prod_{i}n_{i}^{n_{i}/2}}
{(nv)^{n/2}}(1-\frac{nv}{2}\sum_{i}Y_{i}^2) \nonumber \\
&&\approx
\frac{\prod_{i}n_{i}^{n_{i}/2}}
{(nv)^{n/2}}\exp (-\frac{nv}{2}\sum_{i}Y_{i}^2) \nonumber \\
&& \approx
\frac{\sqrt{2\pi}\prod_{i}n_{i}^{n_{i}/2}}
{(nv)^{\frac{n+1}{2}}\sqrt{\sum_{i}\left[ \frac{d\mid \Phi_{\varphi }({\bf r}_{i},t)\mid}{d\varphi}\right]^{2}_{\varphi 
=\tilde{\varphi}}
}
}\delta (\varphi -\tilde{\varphi}).\nonumber
\end{eqnarray}
Finally, averaging over many shots eliminates the symmetry-breaking effect, but in general it
does not flatten out all the ripples since those related to
the creation of collective excitations (an effect due only to interactions) will survive
(see also the model below); thus there is no contradiction with the findings of
\cite{wrong}.

{\it On-chip condensates} As a toy model in which the calculations can be done analytically,
we consider  the case of two condensates launched in the lowest modes $\phi_{l}$
and $\phi_{r}$ of two on-chip atomic waveguides (see e.g. Fig. 1 in \cite{physreva}).
In this case, it is possible to make an (adiabatic) two-level approximation and the problem can be solved 
analytically. In the conditions specified in \cite{physreva}
only the modes $\phi_{1}= (\phi_{l} + \phi_{r})/2$ and $\phi_{2} = (\phi_{l} - \phi_{r})/2$ (corresponding to two 
detection channels similar to \cite{ycastin}) participate in the dynamics.

Consider now what happens after the system (initially in a fragmented state) has evolved for some time; the complex 
amplitude probabilities corresponding to each mode are $A_{1,2}$ of modulus $c_{1} = \sqrt{(1-y)/2}$, $c_{2} = 
\sqrt{(1+y)/2}$ where $y$ is a parameter that depends on time, on the initial phase difference $\varphi$, and on 
the nonlinearity (see \cite{physreva} for explicit notations).
Using the notations of \cite{physreva}, we now imagine, in the spirit of \cite{ycastin}, that we make $k_1$ 
detections in the $\phi_1$ channel
and $k_2$ detections in the $\phi_2$ channel: the resulting (unnormalized) many-body function becomes
\begin{equation}
\int_{0}^{2\pi}d\varphi A_{1}^{k_1}A_{2}^{k_2}\mid \Phi^{(\varphi)}(t) \rangle_{N-k}.
\end{equation}

The maximum value of the quantity $c_{1}^{k_1}c_{2}^{k_2}$ is reached for
$\tilde{y} = (k_{2}-k_{1})/k$, where $k = k_{1}+k_{2}$;  expanding around the maximum value, we get 
$c_{1}^{k_1}c_{2}^{k_2}\approx\left(k_{1}/k)\right)^{k_{1}/2}\left(k_{2}/k\right)^{k_{2}/2
}\exp\left[-(k^3/16k_{1}k_{2})(y-\tilde{y})^2\right]$.

After a large enough number of measurements, $k \gg 1$, the value of $y$ tends to localize
around $\tilde{y}$. Indeed, the worst-case scenario for the localization of $y$ corresponds to a minimum value for the 
factor $k^3 /k_{1}k_{2}$, which (with the restriction $k_{1} + k_{2} = k$), is $4k$, therefore increasing with the 
number of measurements. We then find
\begin{equation}
c_{1}^{k_1}c_{2}^{k_2}\approx 
4\sqrt{\frac{\pi}{k}}\left(\frac{k_1}{k}\right)^{\frac{k_{1}+1}{2}}\left(\frac{k_2}{k}
\right)^{\frac{k_{2}+1}{2}}\delta (y - \tilde{y}).
\end{equation}

Now, if one wants to average over many single-shot measurements, the average number of particles detected
say in channel 1 will be
$N \int_{0}^{2\pi}(d\varphi /2\pi)c^{2}_{1}\neq N/2$
which is the two-mode analogue of a density ripple due to interactions ({\it e.g.} for asymptotically large times $t \rightarrow 
\infty$, this average is $N$, while the corresponding average for $c_{2}^{2}$ is zero; in the noniteracting limit 
both averages would be $N/2$).

Also, in the particular case of zero interaction, we recover the same results as in \cite{ycastin}, using the 
substitution $y=\cos\varphi$, namely localization of the phase near
$\tilde{\varphi} \in \{\arccos\tilde{y}, 2\pi - \arccos\tilde{y}\}$,
\begin{equation}
c_{1}^{k_1}c_{2}^{k_2} \approx 2\sqrt{\frac{\pi}{k}}\left(\frac{k_1}{k}\right)^{\frac{k_{1}}{2}}\left(\frac{k_2}{k}
\right)^{\frac{k_{2}}{2}}\delta (\varphi - \tilde{\varphi}). \nonumber
\end{equation}

{\it Quasi one-dimensional condensates} In quasi one-dimensional infinite uniform gases, the GP equation  has two 
solitonic solutions propagating in the $+z$ and $-z$ directions respectively with speed 
$v_{s}=\sqrt{2\mu}\cos\varphi /2$.
Numerical simulations \cite{bongs,solitons} show that, when recombining two such quasi-condensates with a previously established phase 
difference between them, a train of solitons is in fact formed. Consequently, when two fragmented quasi-condensates interfere, our
theory predicts the appearance of solitonic trains propagating (randomly) in the $\pm$ directions.


To get a better understanding of the left- right- localization process, let us simplify drastically the real 
physical situation and consider that the initial density of the two components is such that, upon connection, only 
two solitons with notches at  $z_{0}=\sqrt{2\mu}t\cos\varphi /2 >0$ and $-z_{0}=\sqrt{2\mu}t\cos(\pi-\varphi 
/2)<0$,  corresponding to phase differences of $0<\varphi <\pi$ and $2\pi -\varphi$, are preferentially formed (for
other phases the propagating ripples will be too small to be measurable).
Then the relevant part of the many-body wavefunction at $t$, when the detection is performed, has the generic form 
of a Schr\"odinger cat state,
\begin{equation}
\mid \Psi \rangle = \mid \Phi_{+} \rangle_{N} + \mid \Phi_{-} \rangle_{N}.
\end{equation}
We show below  that the symmetry $\varphi\rightarrow 2\pi - \varphi$ of this state is broken by the measurement 
process. Absorption processes resulting in atoms detected far enough from the notches (where the wavefunctions 
$\Phi_{\pm}$ differ only by phase factors) do not change significantly the probabilistic weight of the states $\mid 
\Phi_{+} \rangle$ and $\mid \Phi_{+} \rangle$, therefore they do not discriminate between the two solitons. Without 
loss of generality and to simplify the proof, we will discuss only processes involving detection of atoms in either 
of the two notches at $\pm z_{0}$. Consider a sequence of detections, say $\{(+z_{0}), (-z_{0}), (-z_{0}) ....\}$, 
and define $f_{s}$ as the sign of the coordinate of the $s$ atom detected.
The probability ${\cal P}_{\{ f \}}$ of a sequence $\{ f \}=\{f_{1},f_{2},f_{3}, ....,f_{n}\}$ can be calculated by
applying $N$ times the operator $\hat{\psi}(z)$; we find
\begin{equation}
{\cal P}_{\{ f \} } = \prod_{k=2}^{n}\frac{\mid 
\prod_{s=1}^{k-1}\Phi_{-f_{k}}(f_{s}z_{0})\mid^2}{2\sum_{a=\pm}\mid\prod_{s=1}^{k-1}\Phi_{a}(f_{s}z_{0})\mid^2}. 
\label{uhu}
\end{equation}
The sum of probabilities over all possible $2^{N}$ sequences is $\sum_{f}{\cal P}_{f}=1$.
Eq. (\ref{uhu}) has the mathematical property that (for large $N$) ${\cal P}_{f}$ flattens to zero for sequences 
which contain comparable amounts of "+" and "-" detections, so it tends to favor one or the other of the states  
$\mid \Phi_{\pm} \rangle$. In other words, the final result of an experiment will be a distribution
which will reflect the shape of either one of these states.
This can be seen immediately for the particular case of approximately dark solitons, $\Phi_{+}(z_{0})=0$, 
$\Phi_{-}(-z_{0})=0$, for which a single detection event is enough to discriminate between the left- and right- 
propagating solitons. Indeed, after the first detection, the resulting (unnormalized) manybody wavefunction is 
$\hat{\psi}(z_{0})\mid\Psi\rangle = \sqrt{N}[\Phi_{+} (z_{0})\mid \Phi_{+} \rangle_{N-1} + \Phi_{-} (z_{0})\mid
\Phi_{-} \rangle_{N-1}] \approx \mid \Phi_{-} \rangle_{N-1}$ if the atom is detected at $z_{0}$, and
$\hat{\psi}(-z_{0})\mid\Psi\rangle\approx \sqrt{N}[\Phi_{+} (-z_{0})\mid \Phi_{+} \rangle_{N-1} + \Phi_{-} 
(-z_{0})\mid \Phi_{-} \rangle_{N-1}] \approx \mid \Phi_{+} \rangle_{N-1}$ if the atom is detected at $-z_{0}$.
In the general case, our claim can be shown by mathematical induction. Given a sequence $f$ of $n$ measurements, 
out of which $n_{+}$ have been detections at $+z_{0}$ and $n_{-}$ at $-z_{0}$, we can associate, according to Eq.(\ref{uhu}),
to the $n+1$ detection event the probabilities
\begin{equation}
{\cal P}_{\{f,\pm \}} = {\cal P}_{\{ f \} }\left[ 1+\xi^{\pm2(n_{+}-n_{-})}\right]^{-1},
\end{equation}
where $\xi = \mid\Phi_{-}(-z_{0})/\Phi_{-}(z_{0})\mid = \Phi_{-}(-z_{0})/\Phi_{-}(z_{0})\mid <1$. Then ${\cal 
P}_{\{f,+ \}}>{\cal P}_{\{f,- \}}$
if $n_{+}>n_{-}$ and ${\cal P}_{\{f,+ \}}<{\cal P}_{\{f,- \}}$
if $n_{+}<n_{-}$. This shows that the initial difference between $n_{+}$ and $n_{-}$ will increase exponentially 
fast under subsequent detections, which proves our point. As an example, an alternating sequence $\{+, -, +, -, 
..., -\}$ of $n$=even detections will have a probability
$2^{-n/2}\xi^{n}(1+\xi^{2})^{-n/2}$,
while for a constant sequence $\{+, +, +, ..., +\}$ the corresponding probablility is 
$\prod_{k=1}^{n}[1+\xi^{2(k-1)}]^{-1}$; one can immediately check that the ratio between the first probability and 
the
second goes indeed very fast to zero as the number of detections $n$ increases.

In conclusion, we have shown that the evolution of a fragmented-state BEC
yields a many-body state which is a continuous superposition of GP states, embedding the effects of interaction
and indistinguishability. An analysis of the measurement process reveals that this many-body state collapses onto
only one of the GP states in the superposition.

 This work was supported by the Academy of Finland (Projects No. 00857, No. 7111994, and No. 7118122). I 
am grateful to Y.~Castin and A.~J.~Leggett for illuminating discussions.


\begin{thebibliography}{99}

\bibitem{anderson}P.~W.~Anderson, in {\em The Lesson of Quantum Theory},
J.~D.~Boer, E.~Dal, O.~Ulfbeck, Eds. (Elsevier, Amsterdam, 1986), pp. 23-33.

\bibitem{andrews} M.~R.~Andrews, C.~G.~Townsend, H.-J.~Miesner, D.~S.~
Durfee, D.~M.~Kurn, and W.~Ketterle, Science {\bf 275}, 637 (1997).

\bibitem{walls}  W.~Hoston and L.~You, Phys.~Rev.~A {\bf 53}, 4254 (1996); H.~Wallis, A.~Rohrl, M.~Naraschewski, 
and A.~Schenzle, Phys.~Rev.~A {\bf 55}, 2109 (1997); A.~Rohrl, M.~Naraschewski, A.~Schenzle, H.~Wallis, Phys.~Rev.~ 
Lett. {\bf 78}, 4143 (1997).

\bibitem{javanainen} J. ~Javanainen and S.~M.~Yoo, Phys.~Rev.~Lett.~{\bf 76}, 161 (1996).

\bibitem{ycastin}  M.~Naraschewski, H.~Wallis, A.~Schenzle, J.~I.~Cirac, and
P.~Zoller, Phys.~Rev.~A {\bf 54}, 2185 (1996); Y.~Castin and J.~Dalibard, Phys. Rev. A {\bf 55},
4330 (1997).

\bibitem{dalibard} Z.~Hadzibabic, S.~Stock, B.~Battelier, V.~Bretin, and J. Dalibard, Phys.~Rev.~Lett. {\bf 93}, 
180403 (2004).

\bibitem{ketterlenew} Y.~Shin, C.~Sanner, G.-B.~Jo, T.~A.~Pasquini, M.~Saba, W.~Ketterle, D.~E.~Pritchard, 
M.~Vengalattore, and M.~Prentiss, Phys.~Rev.~A {\bf 72}, 021604(R) (2005); G.-B.~Jo, Y.~Shin, S.~Will, 
T.~A.~Pasquini, M.~Saba, W.~Ketterle, D.~E.~ Pritchard, M.~Vengalattore, and M.~Prentiss, Phys.~Rev.~Lett. {\bf 
98}, 030407 (2007).

\bibitem{pritchard} M.~Saba {\it et. al.}, Science {\bf 307}, 1945 (2005);
Y.~Shin {\it et. al.}, Phys. Rev. Lett. {\bf 95}, 170402 (2005).

\bibitem{papers} A.~J.~Leggett and F.~Sols, Found.~Phys.~{\bf 21}, 353 (1991); A.~J.~Leggett, Rev.~Mod.~Phys.~{\bf 
73}, 307 (2001); G.~S.~Paraoanu, S.~Kohler, F.~Sols, and A.~J.~Leggett, J.~Phys.~B {\bf 34}, 4689
(2001); G.~S.~Paraoanu, Ph.~D. thesis, University of Illinois at Urbana-Champaign, 2001.

\bibitem{stringari} L.~Pitaevskii and S.~Stringari, Phys.~Rev.~Lett.~{\bf 83}, 4237 (1999).

\bibitem{mueller} E.~J.~Mueller, T.-L.~Ho, M.~Ueda, and G.~Baym, Phys.~Rev.~A {\bf 74}, 033612 (2006).

\bibitem{wrong} L.~S.~Cederbaum, A.~I.~Streltsov, Y.~B.~Band, and O.~E.~Alon, Phys.~Rev.~Lett. {\bf 98}, 110405 
(2007); H.~Xiong, S.~Liu, and M.~Zhan, New J.~Phys.~{\bf 8}, 245 (2006); S.~Liu and H.~Xiong, New J.~Phys.~{\bf 9}, 
412 (2007).

\bibitem{fetter} A.~L.~Fetter, Ann.~Phys. {\bf 70}, 67 (1972).

\bibitem{castin-gardiner} C.~W.~Gardiner, Phys. Rev. A {\bf 56}, 1414 (1997);
Y. Castin and R. Dum, Phys. Rev. A {\bf 57}, 3008 (1998).

\bibitem{physreva} J.~A.~Stickney and A.~A.~Zozulya, Phys. Rev. A, {\bf 68}, 013611 (2003).

\bibitem{bongs} S. Burger, K. Bongs, S. Dettmer, W. Ertmer, K. Sengstock, A. Sanpera, G. V. Shlyapnikov, and M. 
Lewenstein, Phys.~Rev.~Lett. {\bf 83}, 5198 (1999).


\bibitem{solitons} W. P. Reinhardt and C. W. Clark, J. Phys. B: At. Mol. Opt. Phys. {\bf 30} (1997), L785; J. 
Denschlang, J. E. Simsarian, D. L. Feder, C. W. Clark, L. A. Collins,
J. Cubizolles, L. Deng, E. W. Hagley, K. Helmerson, W. P. Reinhardt, S. L. Rolston, B. I. Schneider, and W. D. 
Phillips, Science {\bf 287}, 97 (2000); S.~Burger, K.~Bongs, S.~ Dettmer, W.~Ertmer, K.~Sengstock, A.~Sanpera, 
G.~V.~Shlyapnikov, M.~Lewenstein, Phys.~Rev.~Lett.~{\bf 83}, 5198 (1999).



\end{thebibliography}
\end{document}